\documentclass[aps,prl,reprint,superscriptaddress,nobibnotes]{revtex4-2}

\usepackage{graphicx}
\usepackage{dcolumn}
\usepackage{bm}
\usepackage{lineno}
\usepackage{ulem}
\usepackage{soul}
\usepackage{color}
\usepackage{array}
\usepackage{booktabs}
\usepackage{multirow, makecell}
\usepackage{physics}
\usepackage[colorlinks=true, allcolors=blue]{hyperref}%

\begin{document}


\title{Exploring two-dimensional van der Waals heavy-fermion material:\\ 
	Data mining theoretical approach}


\author{Bo Gyu Jang}
\affiliation{Theoretical Division, Los Alamos National Laboratory, Los Alamos, New Mexico 87545, USA}

\author{Changhoon Lee}
\affiliation{Department of Chemistry, Pohang University of Science and Technology, Pohang 37673, Korea}
\affiliation{Max Planck POSTECH Center for Complex Phase of Materials, Pohang University of Science and Technology, Pohang 37673, Korea}

\author{Jian-Xin Zhu}
\email{jxzhu@lanl.gov}
\affiliation{Theoretical Division, Los Alamos National Laboratory, Los Alamos, New Mexico 87545, USA}
\affiliation{Center for Integrated Nanotechnologies, Los Alamos National Laboratory, Los Alamos, New Mexico 87545, USA}

\author{Ji Hoon Shim}
\email{jhshim@postech.ac.kr}
\affiliation{Department of Chemistry, Pohang University of Science and Technology, Pohang 37673, Korea}
\affiliation{Division of Advanced Materials Science, Pohang University of Science and Technology, Pohang 37673, Korea}


\date{\today}

\begin{abstract}
The discovery of two-dimensional (2D) van der Waals (vdW) materials often provides interesting playgrounds to explore novel phenomena. One of the missing components in 2D vdW materials is the intrinsic heavy-fermion systems, which can provide an additional degree of freedom to study quantum critical point (QCP), unconventional superconductivity, and emergent phenomena in vdW heterostructures. Here, we investigate 2D vdW heavy-fermion candidates through the database of experimentally known compounds based on dynamical mean-field theory calculation combined with density functional theory (DFT+DMFT). We have found that the Kondo resonance state of CeSiI does not change upon exfoliation and can be easily controlled by strain and surface doping. Our result indicates that CeSiI is an ideal 2D vdW heavy-fermion material and the quantum critical point can be identified by external perturbations.
\end{abstract}


\maketitle

In recent years, there has been a significant interest in 2D vdW materials due to the realization of magnetism in the 2D limit. Over just the past five years, various 2D vdW magnetic materials, such as CrI$_{3}$~\cite{Huang2017}, Cr$_{2}$Ge$_{2}$Te$_{6}$~\cite{Gong2017}, VSe$_{2}$~\cite{Bonilla2018}, MnSe$_{2}$~\cite{OHara2018}, and Fe$_{x}$GeTe$_{2}$ ($x$=3$\sim$5)~\cite{Kim2018, Seo2020} have been reported. These materials contain 3$d$ magnetic elements in their 2D atomically thin crystals and thus exhibit intrinsic magnetic properties. Compared to bulk 3D materials, the 2D vdW magnetic materials provide additional controllability via strain, surface doping, gating, Moir\'e potential, etc. Therefore, the discovery of 2D vdW magnetic materials has provided an interesting route to explore new physical phenomena and design novel devices. The realization of magnetic 2D vdW materials is reminiscent of the possible existence of heavy-fermion state in 2D materials.
 
The heavy-fermion materials containing rare-earth or actinide ions can be regarded as prototypes of lattice Kondo systems, where the interplay between the Ruderman-Kittel-Kasuya-Yoshida (RKKY) interaction and the Kondo effect determines the ground state. When the RKKY interaction dominates, local moments tend to form long-range magnetic order at low temperatures, usually antiferromagnetic (AFM) phase. 
On the other hand, when the Kondo effect prevails, local magnetic moments are screened by itinerant conduction electrons leading to the paramagnetic (PM) Kondo resonance state (strong hybridization between $f$ electrons and conduction electrons; $f-c$ hybridization). 
Due to the small energy scales of two competing interactions, the ground state can be easily tuned through a quantum phase transition by external perturbations, such as pressure, chemical doping or magnetic field~\cite{Hegger2000, Park2006, Seo2015, Jiao2015, Kuchler2006,Sebastian2009, Knebel2001, Iizuka2012,Yamaoka2014,Yuan2003,Ren2014}. In the vicinity of QCP, anomalous phenomena often occur including unconventional superconductivity and non-Fermi liquid. If heavy-fermion state can be realized in 2D vdW materials, it can provide a new platform to study QCP, magnetism and superconductivity with additional controllability.  
 
There have been several attempts to realize the heavy-fermion state in 2D limit. The dimensionality tuning from 3D to 2D heavy-fermion state was firstly reported by using the epitaxially grown CeIn$_{3}$/LaIn$_{3}$ superlattice~\cite{Shishido2010}. Several other artificial Kondo superlattices have been studied to investigate the interaction between two different phases, such as unconventional superconductivity and magnetic order, in the quasi-2D regime~\cite{Mizukami2011, Shimozawa2014, Goh2012, Naritsuka2018, Naritsuka2019, Naritsuka2021}. Recently, it was reported that the artificial 2D rare-earth free heavy-fermion state can be realized from a 1T/1H-TaS$_{2}$ heterostructure~\cite{Vano2021, Ruan2021}. The Kondo coupling between the localized moment in 1T-TaS$_{2}$ layer and the itinerant electrons in 1H-TaS$_{2}$ layer give rise to a heavy-fermion state. Trilayer twisted graphene was also suggested as vdW platform for the realization of heavy-fermion physics~\cite{Ramires2021}. However, a genuine 2D heavy-fermion state within the monolayer limit has not yet been reported. 

A recent high-throughput computational study proposed 1,036 easily exfoliable compounds from experimentally known compounds~\cite{Mounet2018}. Their calculations using vdW density functional theory were validated against random phase approximation calculation and experimental results in already well-known 2D materials. This large portfolio of materials enables us to investigate the novel 2D vdW exfoliable materials with desired properties. In order to explore the potential 2D vdW heavy-fermion candidates, we considered 32 lanthanide materials containing 4$f$ orbitals, which is the essential ingredient of intrinsic heavy-fermion state. We believe that this is a good starting point to explore new 2D vdW heavy-fermion materials.
 
In this study, we investigate possible 2D vdW heavy-fermion materials based on the reported exfoliable lanthanide materials~\cite{Mounet2018}. By analyzing the electronic structure and the Kondo resonance state, Our result reveals that CeSiI system is the most promising 2D vdW heavy-fermion candidate. The dynamical mean-field theory calculation combined with density functional theory (DFT+DMFT) shows that the electronic properties do not change upon exfoliation and the Kondo resonance state of CeSiI monolayer can be sensitively tuned by strain and surface doping. Based on these findings, we predict the ground state of CeSiI monolayer in response to the external perturbation by comparing it with other well-known Ce-based heavy-fermion materials. 

\begin{figure}
	\includegraphics{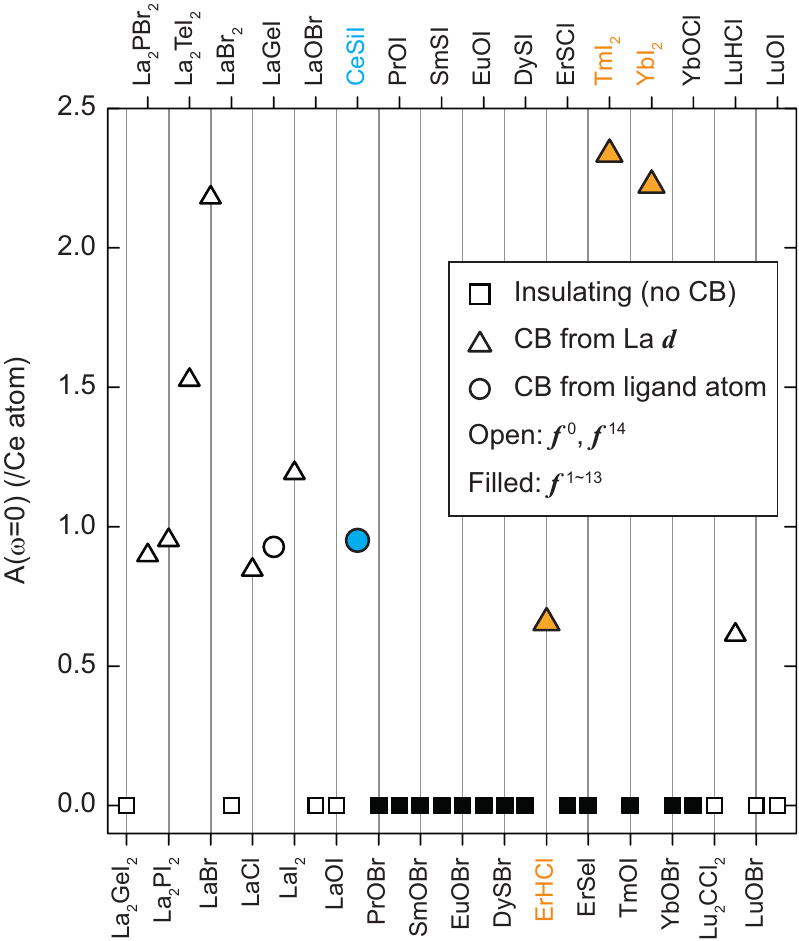}%
	\caption{ Density of states at $E_{F}$ for 32 layered lanthanide materials obtained from DFT calculation within open core method. Materials can be classified into three different groups; insulating (squares), conduction bands (CB) from La d orbitals (triangles), and conduction bands from ligand atoms (circles). Open and filled symbols represent the materials having empty (or fully occupied) 4$f$ orbtials and parially filled 4$f$ orbitals, respectively. } 
	\label{fig:Fig1}
\end{figure}

We first briefly investigate the electronic structures for 32 layered lanthanide materials proposed to be exfoliable. Not only localized $f$ electrons but also itinerant conduction electrons are essential to form heavy-fermion state. Thus, one can easily get an important clue to identify heavy-fermion materials by simply checking the conduction bands at the Fermi level ($E_{F}$). Therefore, we perform DFT open-core calculations, which considers $f$ electrons as core electrons, to examine the existence of non-$f$ conduction bands at $E_{F}$.

Figure~\ref{fig:Fig1} shows the density of states (DOS) at $E_{F}$ for 32 layered lanthanide materials obtained from DFT calculation within open-core approach. Open symbols indicate La ($f^{0}$) and Lu ($f^{14}$) compounds. Since they have empty or fully occupied 4$f$ shells, the lattice Kondo effect cannot be expected. On the other hand, the compounds denoted by square symbols also cannot be heavy-fermion compounds since they exhibit an insulating nature, indicating that there is no itinerant conduction bands to screen localized 4$f$ electrons. Finally, we are left with only four compounds (CeSiI, ErHCl, TmI$_{2}$, and YbI$_{2}$) satisfying the minimal necessary conditions. The four remaining candidates can be again divided into two categories. The conduction bands of CeSiI at $E_{F}$ are mainly coming from the ligand atoms, while those of other compounds are coming from solely lanthanide 5$d$ orbitals.

For comparison, $\alpha$-Ce is a typical heavy-fermion system whose 4$f$ electron is screened by its owns 5$d$ orbitals. In contrast, $\gamma$-Ce, which has an isostructure with a larger volume, shows weak 4$f$-5$d$ hybridization yielding a small Kondo resonance peak~\cite{Kim2019, Jungmann1995}. Since ErHCl, TmI$_{2}$, and YbI$_{2}$ have ligand atoms between their lanthanide atoms, the distance between lanthanide atoms of those compounds are even greater than that of $\gamma$-Ce (3.65). Therefore, 4$f$ electrons are hardly screened by neighboring lanthanide 5$d$ orbitals or the Kondo cloud at each lanthanide site hardly see each other to form the coherent heavy-fermion state. 

Although CeI$_{2}$ is not classified into exfoliable compounds from the previous study~\cite{Mounet2018}, it has the same structure with TmI$_{2}$ and YbI$_{2}$. CeI$_{2}$ has a magnetically ordered ground state and the previous photoemission spectroscopy (PES) study showed that the $f^{1}$ peak is almost absent at $E_{F}$, indicating 4$f$-5$d$ hybridization in CeI$_{2}$ is much smaller than $\alpha$ and $\gamma$-Ce cases~\cite{Jungmann1995}. Our DFT+DMFT calculation on CeI$_{2}$ also verifies the absence of Kondo peak near $E_{F}$ as shown in Fig. S1 of Supporting Information (SI). Considering these aspects, we conclude CeSiI is the most promising candidate for 2D vdW heavy-fermion system among 32 exfoliable lanthanide compounds. Very recently, it was reported that bulk CeSiI has a magnetically ordered ground state~\cite{Okuma2021}. However, it can be sensitively tuned by external perturbations, which will be discussed in the later sections.

\begin{figure*}
	\includegraphics{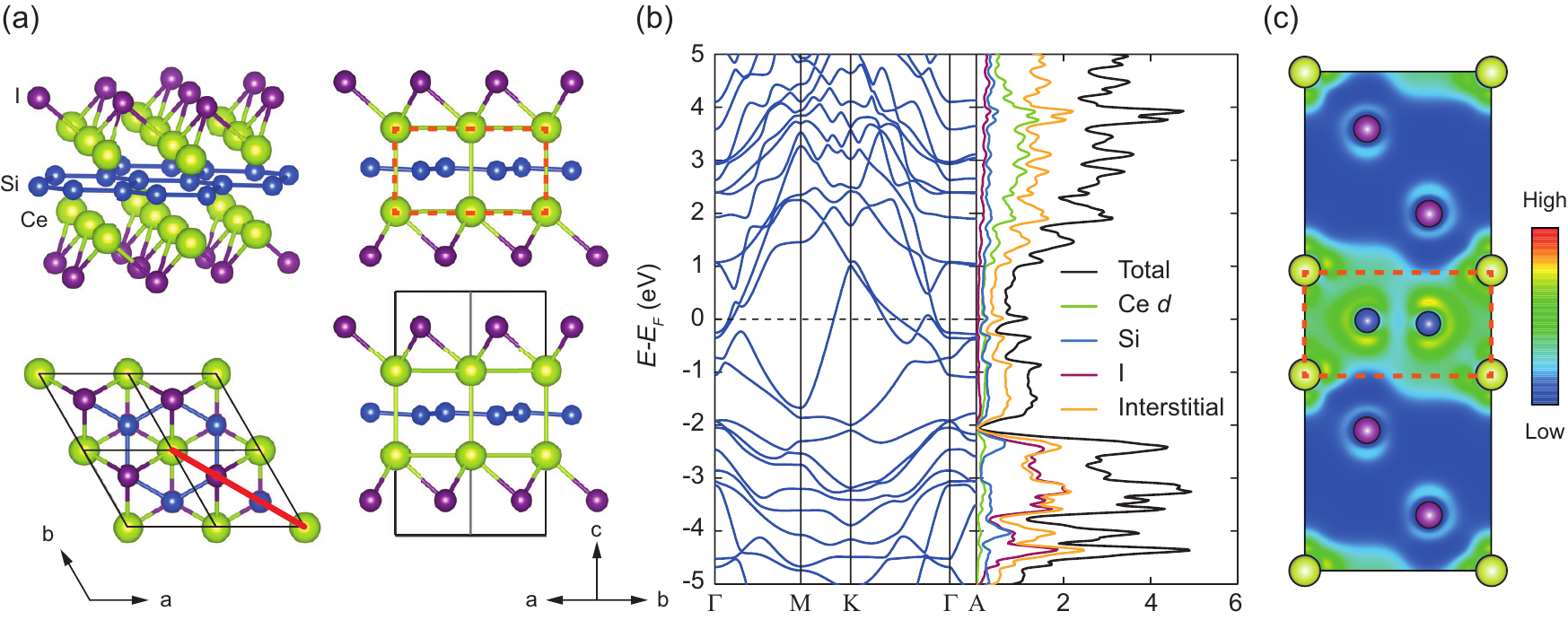}%
	\caption{(a) Crystal structure of CeSiI. Green, blue, and purple spheres indicates Ce, Si, and I atoms, respectively. (b) Band structure and density of state for bulk CeSiI in the absence of 4$f$ electrons. (c) Charge density plot of CeSiI on (110) plane.} 
	\label{fig:Fig2}
\end{figure*}

The CeSiI vdW layer consists of silicene layer, surrounding triangular Ce cage (orange dashed line), and outer iodine layers. The vdW gap can be defined as the iodine to iodine as depicted in the Fig.~\ref{fig:Fig2}(a). The electronic structures calculated by DFT within the open-core approach are presented in Fig.~\ref{fig:Fig2}(b) in terms of DOS (right) and band dispersion (left). Analysis of DOS and band dispersion leads to the following observation: 
(1) The iodine $p$ state occurs primarily between -5 and -2 eV indicating that iodine $p$ state is fully occupied, and their contribution at $E_{F}$ is negligible. 
(2) The band dispersion near $E_{F}$ resembles that of pure silicene compound while the Dirac point at $K$ point is moved to -2 eV in contrast to pure silicene, where the Dirac point is located at $E_{F}$~\cite{Cahangirov2009}. This indicates there is charge transfer from Ce atom to silicene layer.
(3) Therefore, the bands crossing $E_{F}$ should be coming from Si $p_{x}$ and $p_{y}$ antibonding state. However, the contribution from Si at $E_{F}$ is quite small compared to the total DOS as shown in the right panel of Fig.~\ref{fig:Fig2}(b).
 
The most of contribution at $E_{F}$ is originated from the interstitial region. Figure~\ref{fig:Fig2}(c) shows the calculated charge density distribution near $E_{F}$. (energy range -0.15 eV to 0.15 eV with respect to $E_{F}$) for CeSiI on (110) plane (see the red thick line in the left lower panel in Fig.~\ref{fig:Fig2}(a)). The electron density is delocalized all over the Ce cage (between Ce atoms and silicene layer), not localized at specific atoms. This feature agrees well with the Zintl-Klemm feature of [Ce$^{3+}$I$^{-}$Si$^{-}$]$\cdot$$e^{-}$ as suggested by the previous study~\cite{Mattausch1998}. Floating electrons on silicene layer and the interstitial region can be the ingredient of heavy-fermion state. 

\begin{figure}
	\includegraphics{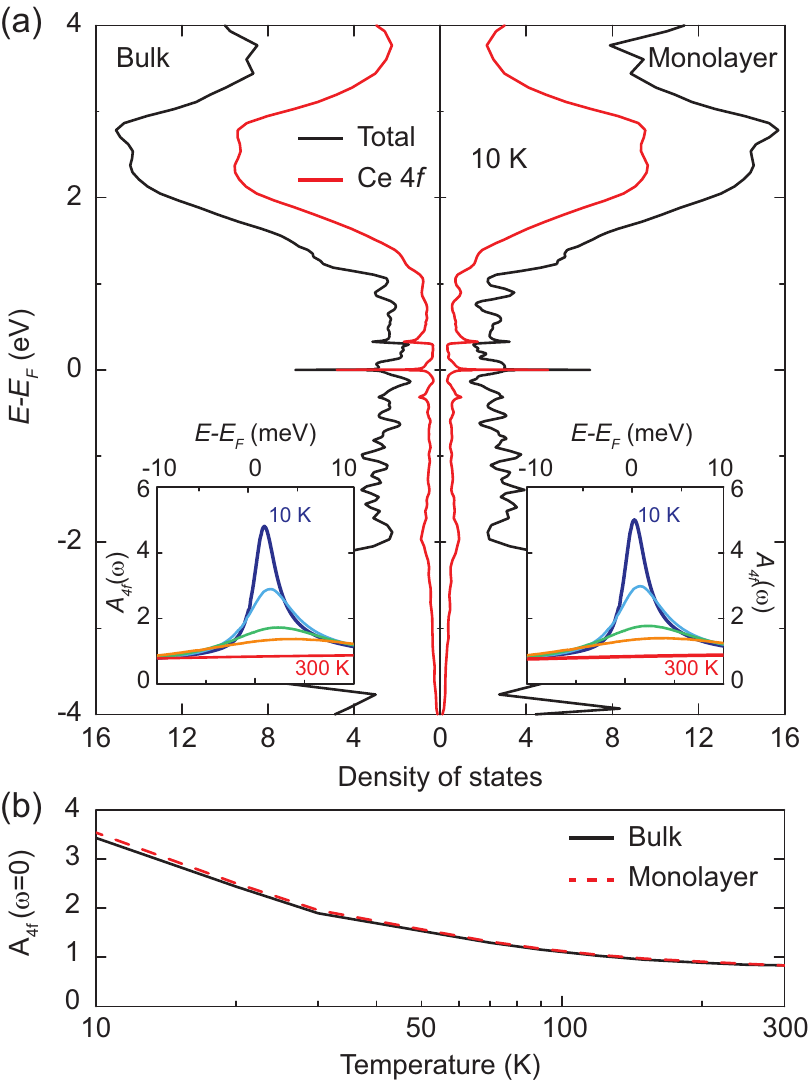}%
	\caption{(a) Electronic structure of bulk and monolayer CeSiI obtained from the DFT+DMFT calculations at 10 K. Inset figures show the Ce 4$f$, $A_{4f}$ evolution as a function of temperature. (b) Temperature evolution of $A_{4f}(\omega=0)$ for bulk and monolayer CeSiI.} 
	\label{fig:Fig3}
\end{figure}

\begin{figure*}
	\includegraphics{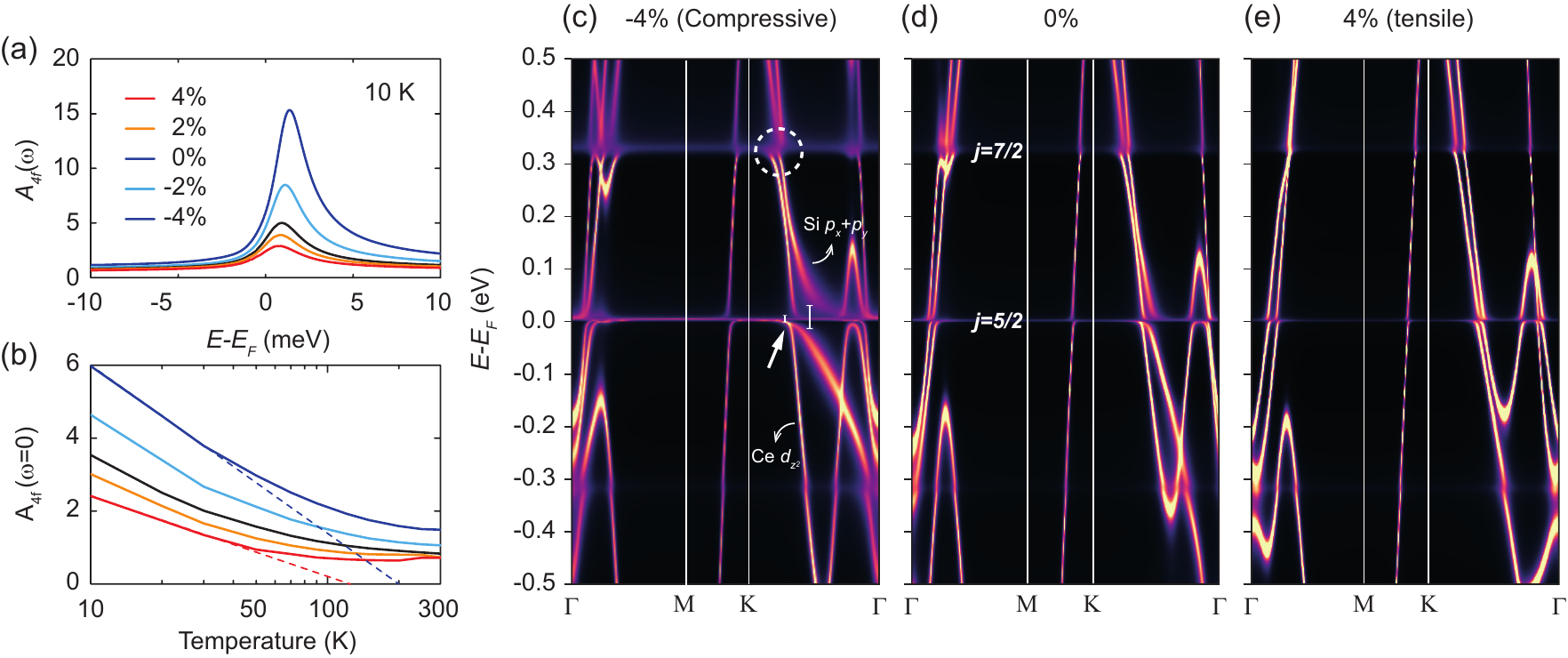}%
	\caption{(a) Ce 4$f$ spectral function, $A_{4f}$ at 10 K depending on strain. (b) Temperature evolution of $A_{4f}(\omega=0)$  depending on strain. (c-e) Momentum–resolved spectral function for   -4\% , 0\%, 4\% strain at 10 K} 
	\label{fig:Fig4}
\end{figure*}

To investigate the hybridization between Ce 4$f$ and conduction electrons ($f-c$ hybridization) properly, DFT+DMFT calculations are performed. The calculated spectral function of bulk and monolayer CeSiI are shown in Fig.~\ref{fig:Fig3}(a) at 10 K. Unlike CeI$_{2}$ case, the Kondo resonance peak is clearly observed near $E_{F}$, indicating substantial hybridization between Ce 4$f$ and conduction electrons. Therefore, CeSiI is located in the vicinity of PM heavy-fermion state.  The electronic structures of bulk and monolayer CeSiI are almost identical, from which similar chemical and physical properties are expected. The temperature-dependent evolution of the Kondo resonance peak also does not change upon exfoliation as shown in the inset of Fig.~\ref{fig:Fig3}(a) and (b). Therefore, CeSiI monolayer can provide an ideal platform for exploring heavy-fermion state in 2D monolayer limit.

Now we investigate the external perturbation effect on the Kondo resonance state of CeSiI monolayer. Figure~\ref{fig:Fig4} shows the electronic structure of CeSiI monolayer with biaxial strain from -4\% to 4\%. Upon compressive (tensile) strain, the volume of Ce cage (orange dashed line in Fig.~\ref{fig:Fig2}(a, c)) surrounding silicene layer decreases (increases). Therefore, one can easily expect stronger $f-c$ hybridization under compressive strain due to the enhanced conduction electron density in the Ce cage. Figure~\ref{fig:Fig4}(a) shows the spectral function of Ce 4$f$ state at 10 K depending on the strain. The Kondo resonance peak is sensitively affected by the strain 

The $T$-dependent 4$f$ spectral function at $E_{F}$, $A_{4f}$($\omega$=0) is shown in Fig.~\ref{fig:Fig4}(b). The Kondo resonance peak is enhanced by the compressive strain for a wide temperature range, resulting in a higher Kondo temperature, $T_{K}$ energy scale. $T_{K}$ can be estimated from the logarithmic $T$ dependence behavior of $A_{4f}$($\omega$=0) $\sim$ ln($T_{K}/T$)  (dashed line in Fig.~\ref{fig:Fig4}(b))~\cite{Choi2012, Kang2019}. The estimated $T_{K}$  for monolayer without strain is 120 K. $T_{K}$ slightly decreases to 110 K upon 4\% tensile strain while it is strongly enhanced up to 200 K upon the compressive strain. The strain-induced $f-c$ hybridization change is also clearly observed in the momentum-resolved spectral function as shown in Fig.~\ref{fig:Fig4}(c-e). Not only the Kondo resonance state from $j$=5/2 state at $E_{F}$ but also those from $j$=7/2 state at 0.3 eV are clearly enhanced upon the compressive strain. The kink feature at 0.3 eV above $E_{F}$ associated with $j=7/2-c$ hybridization is very weak for 4\% tensile strain but it becomes much clear and stronger for compressive strain (dashed circle in Fig.~\ref{fig:Fig4}(c)). In general, the kink feature in the spectral function is related to the formation of the Kondo resonance and the $f-c$ hybridization in heavy-fermion systems~\cite{Choi2013}.

Another notable aspect is the band-dependent $f-c$ hybridization. This feature is most clearly observed in the case of $-4\%$ compressive strain. At the one-third point along $K-\Gamma$ line (white arrow), two dispersive bands are hybridized with Ce 4$f$ states. However, the hybridization strengths of the two bands are quite different. One band is strongly hybridized with Ce 4$f$ state resulting in a large hybridization gap of 50 meV while the hybridization gap from the other band is only 10 meV. This difference arises from the different orbital character of the conduction band. The left conduction band mainly comes from Ce $d_{z^{2}}$, whereas the right band mainly comes from Si $p_{x}$+$p_{y}$ orbitals (See Figure S2 of SI). The hybridization between Ce 4$f$ and Si $p_{x}$+$p_{y}$ states is much stronger and sensitively affected by the strain compared to the hybridization between Ce 4$f$ and Ce $d_{z^{2}}$ state.

\begin{figure}
	\includegraphics{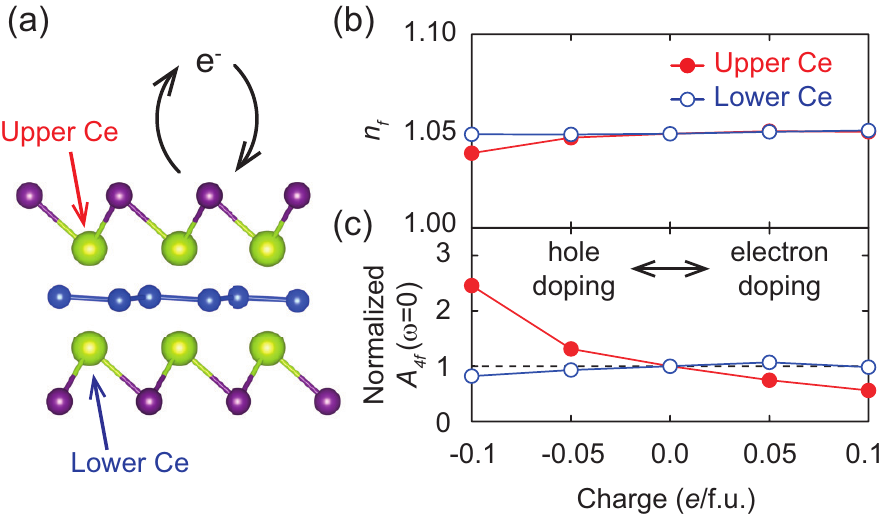}%
	\caption{(a) Schematic picture for surface doping. (b) Ce 4$f$ occupancy of upper and lower Ce layer depending on surface doping. (c) $A_{4f}(\omega=0)$  of upper and lower Ce layer depending on surface doping. They are normalized by pure CeSiI monolayer value.} 
	\label{fig:Fig5}
\end{figure}

Next, we investigate the surface doping effect on the Kondo resonance state. There are two Ce-I layers above and below silicene layer within a monolayer as shown in Fig.~\ref{fig:Fig5}(a). By using the virtual crystal approximation, we control the number of electrons of the upper iodine layer, mimicking the surface doping situation. The calculated Ce 4$f$ orbitals occupancy of each upper and lower Ce layer is presented in Fig.~\ref{fig:Fig5}(b). Regardless of surface doping, Ce 4$f$ electron occupancy remains constant for both Ce layers. However, the Kondo resonance peak of the upper Ce layer is sensitively affected by surface doping while that of the lower Ce layer does not change as shown in Fig.~\ref{fig:Fig5}(c). The surface hole doping significantly enhances only the Kondo resonance state of the upper Ce layer. This suggests that by adjusting surface doping, the Kondo resonance peaks of the upper and lower layer are separately controlled within the monolayer. In addition, I substitution by Te, CeSiI$_{1-x}$Te$_{x}$, would be considered as an effective way to enhance the Kondo resonance of CeSiI monolayer.

\begin{figure}
	\includegraphics{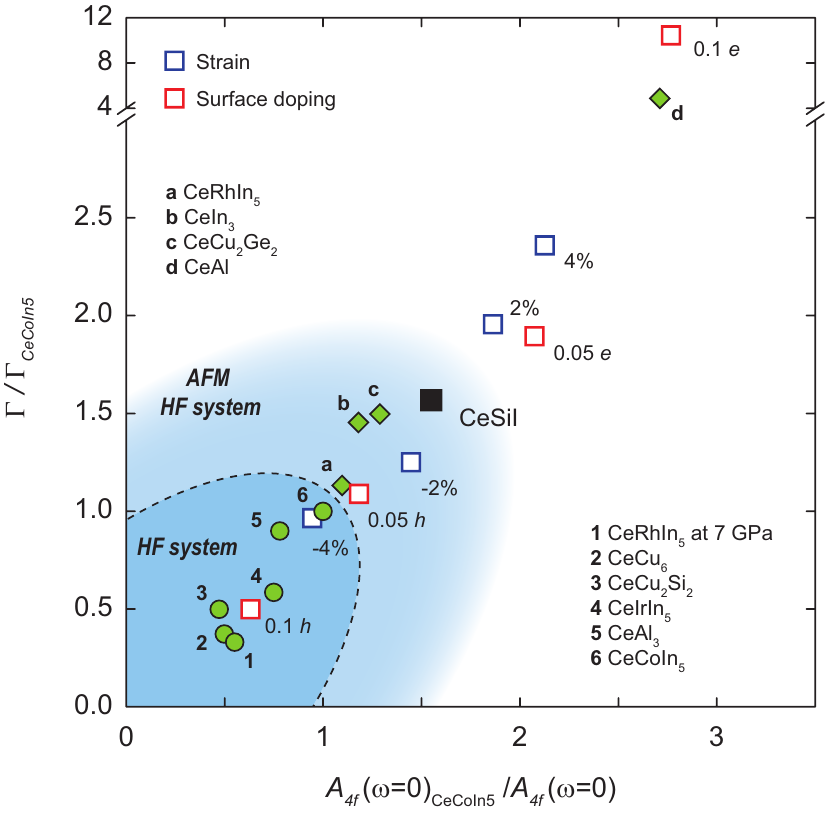}%
	\caption{Phase diagram of Ce-based materials. Inverse density of state at $E_{F}$ ($x$-axis) and inverse quasiparticle lifetime $\Gamma$ ($y$-axis) of CeSiI in comparsion with well-known heavy-fermion systems. Both quantities are normalized by the CeCoIn$_{5}$ values.} 
	\label{fig:Fig6}
\end{figure}

An important remaining question is how sensitive the ground state of CeSiI monolayer can be controlled by strain or surface doping. 
To answer this question, we construct a phase diagram as shown in Fig.~\ref{fig:Fig6}, by calculating the inverse DOS (1/$A_{4f}$($\omega$=0)) at $E_{F}$ ($x$-axis) and inverse quasiparticle lifetime $\Gamma=-Z$Im$\Sigma(0)$ (where $Z^{-1}= m^{*}/m =1-\partial$Re$\Sigma(\omega)/\partial\omega|_{\omega\rightarrow0}$) ($y$-axis) based on PM DFT+DMFT calculations (at $T \sim$ 46 K) and in comparison with well-known Ce-based heavy-fermion materials as shown in Fig.~\ref{fig:Fig6}. Both quantities of each material are normalized by those of CeCoIn$_{5}$ for better comparison. The materials having PM heavy-fermion ground state are denoted by circles while those having magnetically ordered ground states are denoted by diamonds, respectively. Note that CeCoIn$_{5}$ is located in the close vicinity of the QCP and magnetically ordered CeRhIn$_{5}$ undergoes a superconducting transition only above 2 GPa, indicating the proximity to the QCP.  Interestingly, the well-known Ce-based materials are well classiﬁed with respect to the CeCoIn$_{5}$ point. The typical PM heavy-fermion systems are located below CeCoIn$_{5}$ point and magnetically ordered systems are located above the CeCoIn$_{5}$. The PM ground state of CeRhIn$_{5}$ under 7 GPa is also well captured~\cite{Kumar2005, Park2006, Seo2015}.

Along with the well-known Ce-based compounds, CeSiI monolayer with and without external perturbation (strain, surface doping, and Si substitution) are also displayed with black and open squares in Fig.~\ref{fig:Fig6}. For the surface doping case, only the $\Gamma$  and $1/A_{4f}(\omega=0)$ of upper layer Ce are shown in the ﬁgure because those of lower layer Ce do not change upon surface doping on the upper iodine layer. CeSiI monolayer is located above CeCoIn$_{5}$ point, indicating magnetically ordered ground state and it is consistent with the recent experimental result on bulk CeSiI~\cite{Okuma2021}. It is rather located closer to CeIn$_{3}$ and CeCu$_{2}$Ge$_{2}$. Although CeRhIn$_{5}$, CeIn$_{3}$, and CeCu$_{2}$Ge$_{2}$ have magnetically ordered ground states, they have been intensively considered as AFM heavy-fermion systems since their ground state can be easily turned into PM heavy-fermion states by external pressure or chemical doping~\cite{Hegger2000, Park2006, Seo2015, Jiao2015, Kuchler2006,Sebastian2009, Knebel2001, Iizuka2012,Yamaoka2014,Yuan2003,Ren2014}. Therefore, CeSiI should also be located not far from the QCP, and its ground state can be easily tuned by external perturbations like CeIn$_{3}$ and CeCu$_{2}$Ge$_{2}$.

As we have discussed above, compressive strain and surface hole doping strongly enhance the Kondo resonance peak of CeSiI monolayer. With 4\% compressive strain, it is located right below the CeCoIn$_{5}$ indicating possible quantum phase transition to PM heavy-fermion state. With 0.1 surface hole doping, it moves further below CeCoIn$_{5}$ and is located near the well-known PM heavy-fermion materials CeIrIn$_{5}$ and CeCu$_{2}$Si$_{2}$. Therefore, CeSiI monolayer is a promising 2D vdW heavy-fermion candidate and its ground state can be tuned by strain or surface doping (or I substitution by Te), providing alternative ways to explore the QCP, magnetism and unconventional superconductivity. 

The interesting points of this new type of heavy-fermion system are low-dimensionality and frustration originating from its own triangular lattice. During the last couple of decades, the degree of quantum ﬂuctuation has been studied in addition to the Kondo coupling and RRKY interaction. This quantum ﬂuctuation can be enhanced by both lattice frustration and geometrical low-dimensionality. Together with the Kondo coupling strength, this new microscopic quantity (i.e., frustration) enable a construction of a 2D parameter space, zero-temperature global phase diagram~\cite{Si2006, Si2010}. So far, a complete understanding of this global phase diagram is still lacking. 

By enhancing spatial dimensionality to reduce the quantum ﬂuctuation, the cubic Ce$_{3}$Pd$_{20}$Si$_{6}$ was studied~\cite{Custers2012}. One can tune Ce$_{3}$Pd$_{20}$Si$_{6}$ from small Fermi surface (FS) AFM state, through large FS AFM phase, into large FS PM state. Recently, it is reported that geometrically frustrated CePdAl hosts a stable quantum critical phase, possilbly PM small FS phase. YbRh$_{2}$Si$_{2}$~\cite{Friedemann2009} and $\beta$-YbAlB$_{4}$~\cite{Tomita2015} also show similar quantum phase transitions. Although they do not have geometrical frustration, their quasi-2D crystal structure may enhance the quantum ﬂuctuation and lead to a similar quantum phase transition trajectory on the global phase diagram.

However, understanding of the microscopic processes remains missing. CeSiI monolayer can provide a new platform to study the global phase diagram. Due to its low dimensionality and lattice frustration, one can study the upper part of the global phase diagram, where the quantum ﬂuctuation is strong. Besides, this 2D vdW heavy-fermion system has additional control parameters. In addition to magnetic ﬁeld and pressure (uniform strain), uniaxial strain can be an external tuning parameter in CeSiI monolayer. By inducing anisotropy via uniaxial strain, frustration can be controlled, providing a unique opportunity to explore the entire global phase diagram.

Another interesting aspect is the critical behavior of resistivity near this QCP with low-dimensionality or frustration. In the epitaxially grown CeIn$_{3}$/LaIn$_{3}$ superlattice, the quantum phase transition within 2D limit was observed by reducing the superlattice period. Near the QCP, the resistivity shows $T$-linear behavior deviating from the Fermi liquid behavior $\rho$($T$)= $\rho_{0}$ + $AT^{\alpha}$, where $\alpha$=2. This linear $T$-dependence ($\alpha$=1) is also in contrast to the resistivity behavior observed near the pressure-induced QCP in the bulk CeIn$_{3}$ ($\alpha$=1.5)~\cite{Shishido2010}. YbRh$_{2}$Si$_{2}$ also shows $T$-linear resistivity near the QCP~\cite{Friedemann2009, Nguyen2021}, while $\beta$-YbAlB$_{4}$ recovers the Fermi liquid behavior ($\alpha$=2) at very low temperature~\cite{Tomita2015}. Therefore, it would be interesting to study the resistivity behavior in CeSiI monolayer to further understand the strange metallic behavior in the global phase diagram.

Finally, CeSiI monolayer is a new building block to study emergent phenomena in vdW heterostructures. Although 1T/1H-TaS$_{2}$ heterostructure and twisted trilayer graphene can also be 2D heavy-fermion building blocks, delicate preparation is required since those heavy-fermion states are not intrinsic. It would be interesting to study vdW heterostructures which consist of CeSiI monolayer and other 2D vdW components, including magnetic materials. Although the epitaxially grown Kondo superlattice, such as CeCoIn$_{5}$/CeIn$_{3}$ and CeCoIn$_{5}$/CeRhIn$_{5}$, have been used to study the dimensionality eﬀect and the interactions between heavy-fermion state (or superconducting state) and magnetism (bosonic excitations)~\cite{Naritsuka2018, Naritsuka2019}, CeSiI monolayer provides more degrees of freedom to make various interfaces in combination with other 2D vdW materials or substrates.

To summarize, we have investigated 2D vdW heavy-fermion systems from the experimentally known bulk compounds by using DFT+DMFT calculations. Among them, CeSiI is the most promising candidate. We have found that the Kondo coupling strength of CeSiI does not change upon exfoliation and can be easily controlled by strain and surface doping. Finally, we have predicted the ground state of CeSiI monolayer in response to the external perturbations. Although the CeSiI monolayer has a magnetically ordered ground state, we found that the compressive strain and surface hole doping strongly enhance the Kondo coupling strength, indicating possible quantum phase transition to PM heavy-fermion state. Our result thus suggests that CeSiI monolayer can be a genuine 2D vdW heavy-fermion system and provide a new playground to study the QCP and emergent phenomena in vdW heterostructures.

\section{Acknowledgements}
Bo Gyu Jang thanks Junwon Kim and Young-Woo Son for fruitful discussion. 
Work at Los Alamos was carried out under the auspices of the U.S. Department of Energy (DOE) National Nuclear Security Administration (NNSA) under Contract No. 89233218CNA000001. It was supported by UC Laboratory Fees Research Program (Grant Number: FR-20-653926) and in part by Center for Integrated Nanotechnologies, a DOE BES user facility. 
J.H.S and C.L was supported by the National Research Foundation of Korea (NRF) grant funded by the Korea Government (Grant No. 2020R1A5A1019141, No. 2020R1F1A1052898, No. 2020M3H4A2084418, and No. 2021R1F1A1063478). 

\setcounter{table}{0}
\renewcommand{\thetable}{S\arabic{table}}%
\setcounter{figure}{0}
\renewcommand{\thefigure}{S\arabic{figure}}%
\setcounter{equation}{0}
\renewcommand{\theequation}{S\arabic{equation}}%

\section{Supplementary Information }

\subsection{Computational Details}
We first performed DFT open-core calculation using the Vienna \textit{ab-initio} Simulation Package (\texttt{VASP})\cite{Kresse1996}. The experimental crystal structure of 32 lanthanide materials employed from Inorganic Crystal Structure Database (ICSD) were used in the calculations. For structural relaxations of the strained CeSiI cases, DFT open-core calculation with \texttt{VASP} were also employed since DFT calculations with 4$f$ orbitals significantly overestimate $f-c$ hybridization, resulting in smaller lattice constant. (DFT open-core calculation with vdW correction well reproduced the experimental crystal structure of CeSiI.)

In DFT+DMFT calculation, CTQMC solver was basically adopted for all calculations (Fig. 6). The temperature evolution of Kondo resonace peak on the real axis was analyzed by using OCA solver to avoid the ambiguity arose from analytic continuation which is inevitable for CTQMC solver (Fig. 3 and 4). 
The hybridization energy window from -10 to 10 eV with respect to the $E_{F}$ was chosen and $U$ = 5 eV and $J$ = 0.68 eV were used for all compounds. 
For virtual crystal approximation (VCA) calculation, we did not consider the crystal structure change and just used CeSiI monolayer structure.   

\subsection{Spectral function of CeI$_{2}$}

\begin{figure}[h]
	\centering
	\includegraphics[width=8.5cm]{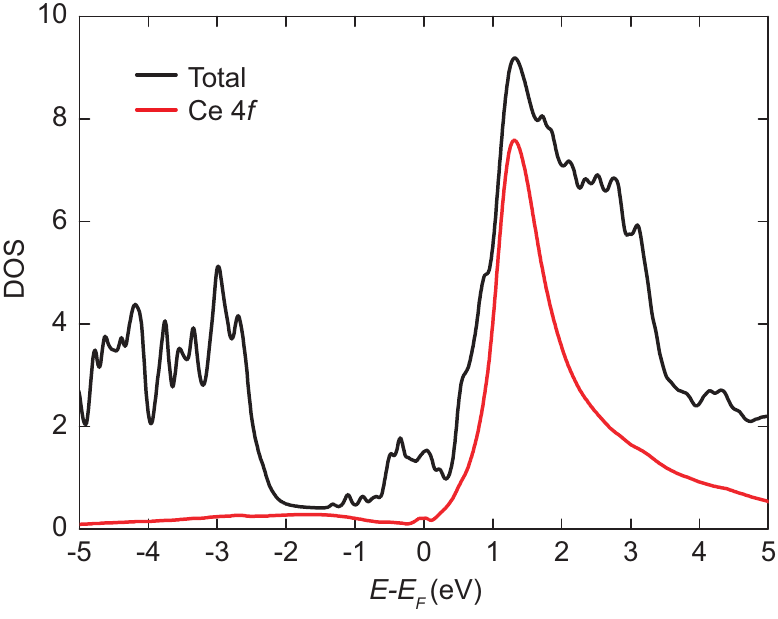}
	\caption{Spectral function of CeI$_{2}$ at 46 K ($\beta$ = 250 eV$^{-1}$) }
	\label{fig:Fig1}
\end{figure}

Figure S1 shows CeI$_{2}$ spectral function obtained from DFT+DMFT calculation with CTQMC solver. The Kondo resonace peak is absent which well agrees with the previous phtoemission spectroscopy (PES) result~\cite{Jungmann1995}.

\subsection{CeSiI DFT band structure with orbital characters}

\begin{figure}[h]
	\centering
	\includegraphics[width=8.5cm]{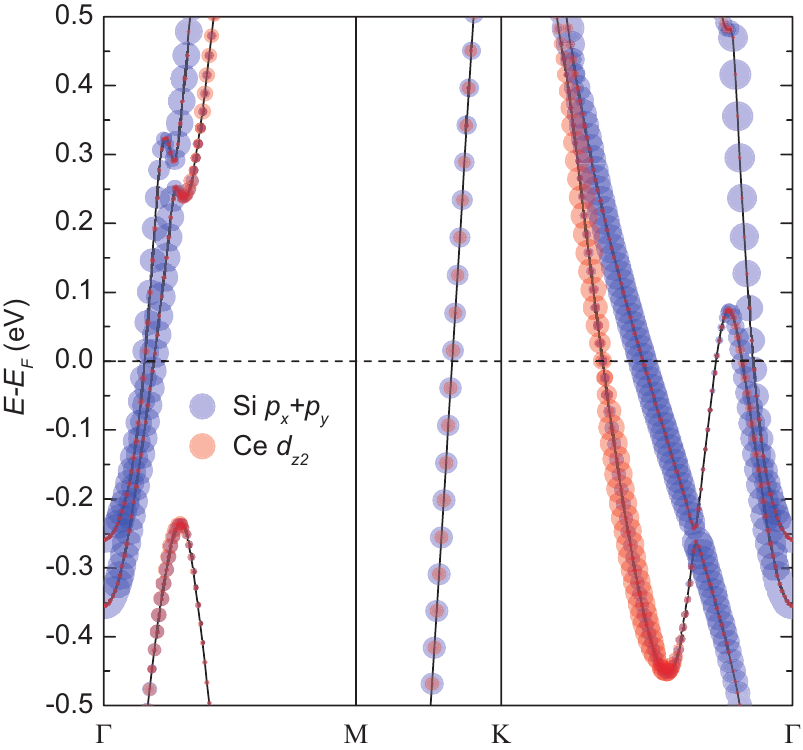}[h]
	\caption{Fat band plot of CeSiI obtained from DFT calculation without 4$f$ orbtials}
	\label{fig:Fig2}
\end{figure}

Figure S2 shows CeSiI fat bands plot obtained from DFT calculation without 4$f$ orbitals. The size of blue and red circles indicate the contribution from Si $p_{x}+p_{y}$ orbitals and Ce $d_{z^{2}}$ orbital, respectively.

\subsection{Si substitution effect on the Kondo resonace state}

\begin{figure*}
	\centering
	\includegraphics{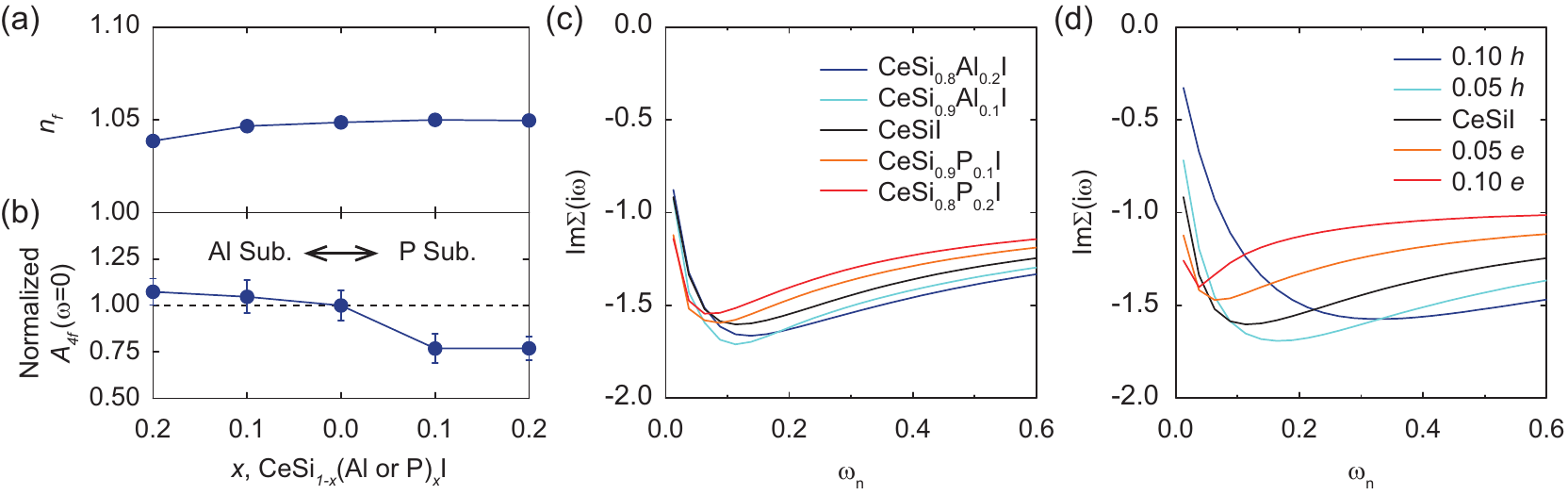}
	\caption{(a) Ce 4$f$ occupancy and (b) $A_{4f}(\omega =0)$ depending Si substitution. Imaginary part of self-energy on Matsubara frequency depending on (c) Si substitution and (d) surface doping.}
	\label{fig:Fig1}
\end{figure*}

The effect of Si substitution is not effecitve like surface doping case. We also adopted the virtual crystal approximation to mimic the partial Si substitution by Al or P atom, CeSi$_{1-x}$Al(P)$_{x}$I. Ce $4f$ orbitals occupancy is not affected by Al or P substitution like the surface doping case as shown in Fig. S3 (a). Figure S3 (b) shows the Ce $4f$ spectral function at the $E_{F}$ as a function of Al or P substitution. The change of Kondo resonance peak depending on substitution is not sensitive like the surface doping case, although the trend is consistent with surface doping case (or I substitution). The error bars indicate the uncertainty arose from the Monte Carlo noise. Considering this noise, the Al substitution effect is almost negligible.  

Figure S3 (c) and (d) present the calculated imaginary part of self-energy depending on Si substitution and the surface doping (or I substitution), respectively. From the calculated self-energy, one can also find that the Si substitution is not effective like the surface doping. The surface hole doping cleary enhance the coherency at low frequency, while the correlation strength increases at higher freqeuncy. Although the high frequency trend in Si substitution cases agree with the surface doping case, the change in low frequency is not clear.

Here, two competing effects are expected in Al and P substitution cases. First effect is the conduction electron density of Ce cage as we discussed in the strain case. Al (P) substitution decreases (increase) the conduction electron density in the Ce cage like the tensile (compressive) strain case. Therefore, it seems that Al (P) substitution suppresses (enhances) the Kondo resonance state. Although we did not consider volume change depending on substitution, volume change could enhance this trend since Al (P) has larger (smaller) ionic radius compared to Si.  From the surface doping result, however, the opposite effect is expected. The shift of conduction bands relative to Ce $f$ bands in Al (P) substitution case is similar to that of surface hole (electron) doping case. Therefore, it is expected that Al (P) substitution enhances (supresses) the Kondo resonance state from dopin point of view. Therefore, it seems that Si substitution by Al or P is not an effective way to control the Kondo resonace state.

\bibliography{CeSiI}

\end{document}